\documentclass[galaxies,article,accept,moreauthors,pdftex,10pt,a4paper]{mdpi}  
\firstpage{1} 
\articlenumber{x}
\doinum{10.3390/------}
\pubvolume{xx}
\pubyear{2016}
\copyrightyear{2016}
\externaleditor{Academic Editor: name}
\history{Received: date; Accepted: date; Published: date}
\pdfoutput=1

\Title{Radiative Transfer Modeling of Radio-band Linear Polarization Observations as a Probe of the Physical Conditions in the Jets of $\gamma$-ray Flaring Blazars }

\Author{Margo  F. Aller $^{1}$, Philip A. Hughes $^{1}$, Hugh D. Aller $^{1}$, Talvikki Hovatta $^{2,3}$, Venkatessh Ramakrishnan $^{2}$}

\AuthorNames{Firstname Lastname, Firstname Lastname and Firstname Lastname}

\address{%
$^{1}$ \quad Department of Astronomy, University of Michigan, Ann Arbor, MI 48109-1107  USA \\
$^{2}$ \quad Aalto Observatory, Mets\"ahovi Radio Observatory, Kylm\"al\"a  Finland         \\
$^{3} $ \quad  Aalto University Department of Radio Science and Engineering, P.O. Box 13000, FI-00076, Aalto, Finland} 
\corres{Correspondence: mfa@umich.edu; Tel.: 001-734-764-3465}

\abstract{Since the mid-1980s the shock-in-jet model has been the preferred paradigm to explain radio-band flaring in blazar jets. We describe our radiative transfer model incorporating relativistically-propagating shocks, and illustrate how the 4.8, 8, and 14.5 GHz linear polarization and total flux density data from the University of Michigan monitoring program, in combination with the model, constrain jet flow conditions and shock attributes. Results from strong  {\it Fermi}-era flares in 4 blazars with widely-ranging properties are presented. Additionally, to investigate jet evolution on decadal times scales we analyze 3 outbursts in OT~081 spanning nearly 3 decades and find intrinsic changes attributable to flow changes at a common spatial  location, or, alternatively, to a change in the jet segment viewed. The model’s success in reproducing these data supports a scenario in which relativistic shocks compress a plasma with an embedded passive, initially-turbulent magnetic field, with additional ordered magnetic field components, one of which may be helical.}


\conferencetitle{Blazars through Sharp Multi-Frequency Eyes}

\begin{document}


\section{Introduction}

The shock-in-jet model has been the preferred paradigm for explaining flaring in blazar jets since the 1980s \cite{MG85, HAA85}. In the radio-band the evidence supporting this scenario is the spectral evolution of the linear polarization during total flux density outbursts, consistent with the expected changes as a propagating shock compresses an initially-turbulent magnetic field; this compression increases the particle density and the magnetic field energy density, and hence the emissivity, and partially orders the magnetic field. The signatures of the passage of a shock in the light curves are a flare in total flux density (S) with a temporally-associated  increase in the fractional linear polarization (P\%) and a systematic swing in the electric vector position angle (denoted by EVPA or $\chi$).  An example long-term light curve which illustrates the shock signatures — temporally well-separated outbursts in S (an envelope over blended, individual flares); outbursts in polarized flux (ranging from a few percent to $\sim$15\%) and ordered changes in the EVPA on timescales of a few years — is shown in Figure 1.

Prompted by renewed interest in models to explain blazar variability with the discovery that $\gamma$-ray flaring from blazars is common \cite{ACK15} we recently revisited the shock-in-jet model. In the new work we allow the shocks to be orientated at any direction to the flow \cite{HAA11} while the original modeling \cite{HAA85} assumed that all shocks were oriented transversely to the flow direction, and we consider the sensitivity of the model fitting to the inclusion of an ordered component of the magnetic field \cite{HAA15}. By comparing time segments of the centimeter-band total flux density and linear polarization data at three centimeter-band frequencies from the University of Michigan blazar variability program (UMRAO) for four $\gamma$-ray-flaring blazars with radiative transfer simulations incorporating propagating shocks, we have identified the jet flow properties and shock attributes \cite{AHA14, AHAJ14}. We summarize those results here, and we use this method to identify long-term changes spanning nearly 3 decades in the flow of a newly-modeled source exhibiting more modest $\gamma$-ray flaring, OT~081, during the first 4 years of {\it Fermi} (see http://www.bu.edu/blazars/VLBA\_GLAST/1749.html). 

\begin{figure}[H]
\centering
\includegraphics[width=3.8in]{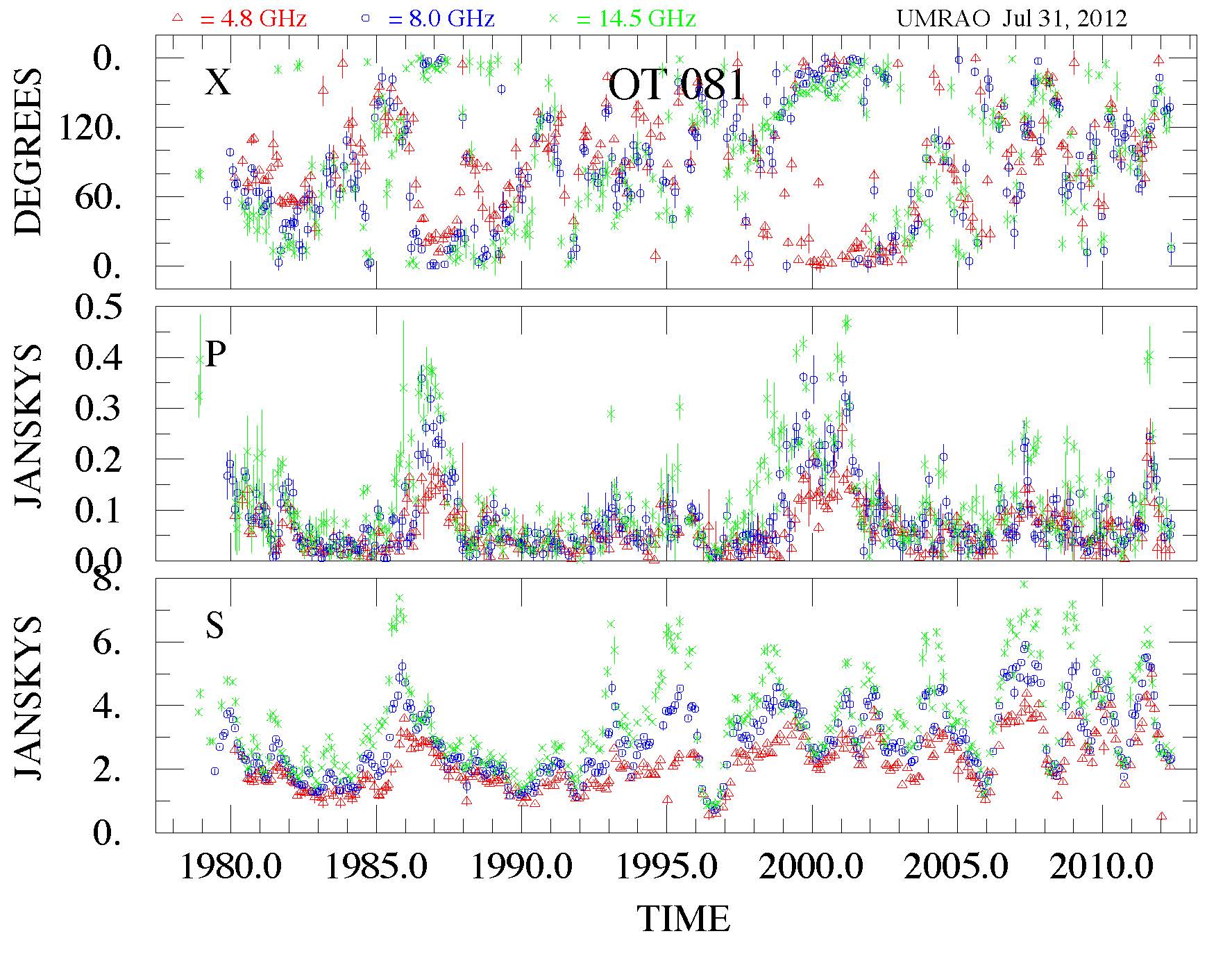}
\caption{ From bottom to top: monthly-averaged total flux density, linear polarization, and electric vector position angle observations for OT~081 (1749+096) from the UMRAO variability program. The three centimeter-band frequencies are symbol and color coded  as denoted at the top left.}
\end{figure}   

\section{The Model and Method}

The fundamental parameters which describe the jet and shock system and their observational constraints, are listed in Table 1. These define the internal state of the quiescent flow, the bulk dynamics of the flow and its orientation, and the shock attributes. VLBA imaging data at 43 GHz from the Boston U. program and at 15 GHz from MOJAVE provide independent constraints during the modeling and checks on the results for some parameters. For example, the component speed, $\beta_{app}$, obtained from these complementary VLBA observations is used to set the shock sense (forward or reverse relative to the quiescent flow).

\begin{table}[H]
\caption{Free Parameters of the Jet \& Shock System and the Constraining Data.}
\small 
\centering
\begin{tabular}{cc}
\toprule
\textbf{Model Parameter}	& \textbf{UMRAO Data Constraint}\\
\midrule
Low energy cutoff ($\gamma_{i}$)   & EVPA Spectral Behavior	\\
Bulk Lorentz Factor ($\gamma_{f}$)  & P\%                        \\
Viewing Angle ($\theta$)           &  P\%                      \\
Axial B Field	(B$_{z}$)          &   EVPA and P\%		\\
Shock obliquity ($\eta$)           & $\Delta$EVPA               \\
Shock compression ($\kappa$)       & $\Delta$S and P\%          \\
Shock onset (t$_0$)                 & Start of flare in S or P  \\
Shock length (l)                   & Duration of flare in S    \\
Shock sense (F or R)               & Doppler factor and $\beta_{app}$  \\
\bottomrule
\end{tabular}
\end{table}

The model assumes that a population of radiating particles with a power law distribution of Lorentz factor with cutoff $\gamma_i$ permeates the quiescent flow. The magnetic field within the jet is assumed to be predominantly turbulent in the radio-emitting region for the reasons summarized in \cite{HUG05}, but the formulation allows for the inclusion of two ordered magnetic field components; one is axial, originally introduced as 2\% of the energy density in the random component to explain the stable EVPAs apparent in the UMRAO data during quiescence \cite{HAA11}, but later increased to tens of percent to provide a better fit to the data \cite{AHA14}. A second ordered magnetic field component included as a free parameter in the model may be helical in character. The shock orientation is permitted to be at an arbitrary orientation to the flow direction and is specified by its obliquity, and by a second angle, the azimuthal direction of the shock normal, which has been shown to have little affect on the simulations \cite{HAA11}.  The shocked flow itself is specified by a length expressed as a percentage of the flow and a compression factor relative to unity. The compression factor is a measure of the shock strength. The shock orientation relative to the flow direction is assumed to be the same for all shocks within the modeled time window. 

Details of the adopted iterative modeling process are given in \cite{AHA14}. The number of shocks and their onsets are set to match the structure in the UMRAO total flux density and linear polarization light curves, the duration is obtained from the combination of the observed structure in the light curves and the expected flare shape based on a library of simulations for a single shock, and the shock obliquity is set by the range of change in the EVPAs. In selecting outbursts for modeling, those exhibiting  the signature of transverse shocks are preferentially selected since changes through large ranges are easier to identify in the, often complex, EVPA light curves; these systematic changes in EVPA are  most apparent in the data at 14.5 GHz where the sampling is densest. 

\section{Results}
\subsection{Single-Epoch Modeling: Study of Jet Properties during the {\it Fermi} Era.}
A motivation for the modeling is to identify flow conditions in the jets of $\gamma$-ray flaring blazars. In the initial phase of this work we analyzed individual time segments of the UMRAO data in 4 sources exhibiting moderate to strong $\gamma$-ray flares. The sources were selected based on the appearance of well-separated outbursts in the UMRAO data, and simultaneous or near-simultaneous flaring in the centimeter and GeV bands. In the case of one source, 1156+295, two adjacent time windows were modeled which included nearly-identical radio-band outbursts in terms of amplitude and spectral evolution in total flux density and maximum linear polarization amplitude. However, the second radio-band outburst was paired with a $\gamma$-ray flare, while the first outburst was an orphan radio outburst (no temporally-associated $\gamma$-ray flare).  The difference in the $\gamma$-ray to radio-band activity  was attributed to the presence of a larger number of shocks in the paired event \cite{HAA15} and more complex structure at VLBI scales found from 43 GHz imaging data \cite{AHAT14}.

The data for the most recently modeled source in this group, 0716+714, are shown in Figure 2 left. MOJAVE source-integrated data taken from the MOJAVE webpage
at http://www.physics.purdue.edu/astro/MOJAVE/sourcepages/0716+714.shtml are included for comparison.  The fact that this blazar is extreme is apparent from the unusual shape of the outbursts in total flux density (lower panel in Figure 2 left). These exhibit an unusual triangular shape with nearly equal rise and fall times, and  several of these events are captured within the time window modeled. Also, the maximum fractional linear polarization is unusually high, near 14\% at 14.5 GHz, compared with the values for the other sources modeled (typically 5\% to 8\% at maximum). The adopted shock onsets used in the simulation shown in the right figure are denoted by upward purple arrows along the abscissa; these onsets correspond to the times at which the leading edge of the shock enters the flow.
\begin{figure}[H]
   \begin{center}$
\begin{array}{cc}
   \includegraphics[width =3.5in]{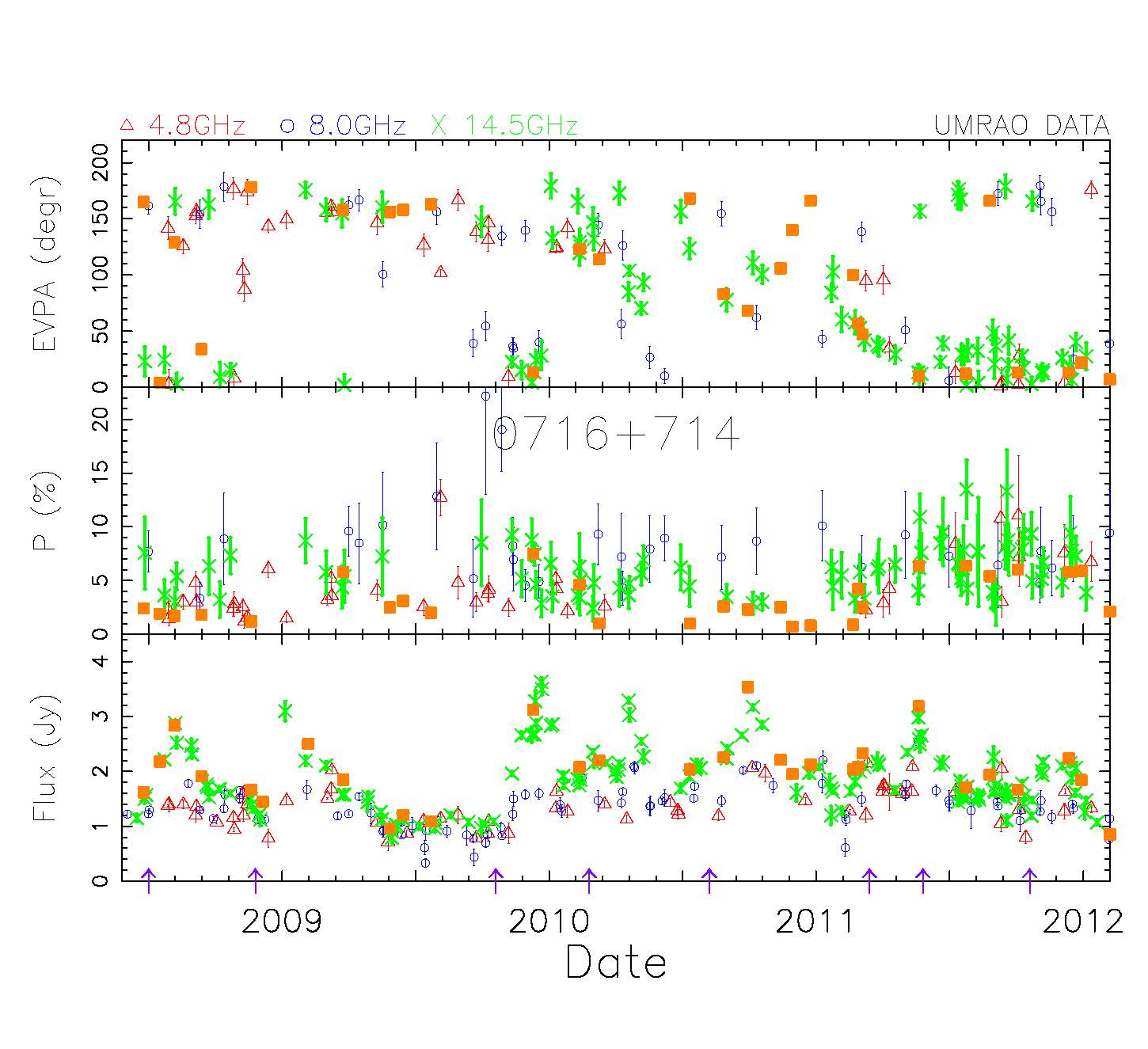} &
\hfill
  \includegraphics[width=3in]{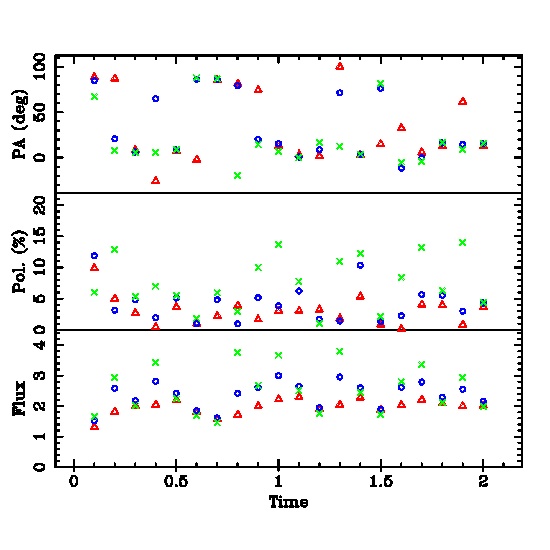}
\end{array}$
\end{center}
\caption{Left: Daily-averaged centimeter-band data for 0716+714 during the modeled time window. UMRAO data are color and symbol coded as in Figure 1. MOJAVE data are shown with gold squares. Right: The simulated light curves based on the adopted jet and shock parameters.}
\end{figure}  

In the simulation shown to the right the three model frequencies are color and symbol coded to match those used for the data. The scaling of time in the simulation is set by the duration of the activity  modeled, while the total flux density amplitude is scaled to match the peak value at the highest UMRAO frequency, 14.5 GHz. This simulation reproduces the general character of the variability including  the spectral behavior as a function of time, the amplitude range of the total flux density flares, and the global shape of the outbursts and the position of features. Note that in this source a multi-year time window has been modeled with a single set of jet parameters. However, while many of the global features are simulated with the adopted model, refinements are required to reproduce the spectral character of the 8 GHz polarization, especially during shocks 2 through 5. In the early part of the simulation the values of P\% are too high at all three frequencies; this is an artifact of the modeling which starts from quiescence and neglects the effect of earlier activity.

The derived jet flow and shock attributes for the 4 blazars modeled in this phase of the work are summarized in Table 2.  As expected, several parameters found for 0716+714 are extreme compared with the other 3 blazars modeled. These include  the high bulk Lorentz factor which was determined from the unusually high value of  P\% and which is consistent with the high values of $\beta_{app}$ determined from VLBI observations for the fastest components by MOJAVE \cite{LIS16} and other programs, e.g. \cite{RAN15}. 
\begin{table}[H]
\caption{Flow Parameters and Shock Attributes from Radiative Transfer Modeling}
\small 
\centering
\begin{tabular}{ccccc}
\toprule
\textbf{Parameter} & \textbf{0420-014}	& \textbf{OJ 287}  &  \textbf{1156+295 flare2}   & \textbf{0716+714}\\
\midrule
Cutoff Lorentz Factor	& 50 			&  10              &   50                 &  50 \\
Bulk Lorentz Factor 	& 5			&     5            &   10                 &  20  \\
Viewing Angle     & 4$^{\circ}$          & 1.5$^{\circ}$   & 2.0$^{\circ}$        & 12$^{\circ}$  \\
Axial B (energy density)           & 16\%                 & 50\%            & 50\%                 & 36\%          \\ 
Shock Sense        &       F            &    F            &  F              &    F                    \\         
Number of Shocks & 3                     & 3               &    4                 &   8  \\
Shock Obliquity  & 90$^{\circ}$          & 30$^{\circ}$    &    90$^{\circ}$      & 90$^{\circ}$ \\
Shock Compression & 0.65 - 0.8          & 0.5 - 0.7      &    0.5 - 0.8        & 0.17 - 0.27 \\
Shock $\beta_{app}$ (in units of c) &  11              & 17            &   22               &  9.5       \\
\bottomrule
\end{tabular}
\end{table}

 Another notable property is the unusually-high viewing angle when compared with the other modeled blazars and with results based on alternative analysis procedures for 0716+714 \cite{LJM13, RAN15}. While a viewing angle of 12$^{\circ}$ provides the best match between the simulation and the data, we could not rule out values as low as 5.8$^{\circ}$ \cite{LJM13} since the model linear polarization is less sensitive to changes in viewing angle at values of 
$\theta\geq$5$^{\circ}$. 

\subsection{Multi-epoch Modeling: internal changes in the jet on timescales of years to decades}
   In order to look for temporal changes in jet flows on time scales of decades to years we have modeled three epochs of the UMRAO data for the highly-compact source OT~081: 1985.0 to 1986.0 (T1985), 2008.4 to 2010.7 (T2008) and 2010.8 to 2011.9 (T2010). As shown in Figure 1 this source exhibited strong and persistent variability throughout the UMRAO program making it an ideal target for a long-term study.  Further, the data used in the analysis presented here were consistently obtained with the 26-m paraboloid under automatic computer control using the same observing and reduction procedures throughout; hence the work is based on a homogeneous data set. The UMRAO data for the 1985 outburst had previously been modeled using an earlier simulation code \cite{HAA91}. This work identified reverse shocks  in order to obtain the slow flows (of order c) expected based on the limited VLBI data available at the time.

\begin{figure}[H]
   \begin{center}$
\begin{array}{cc}
   \includegraphics[width =3.5in]{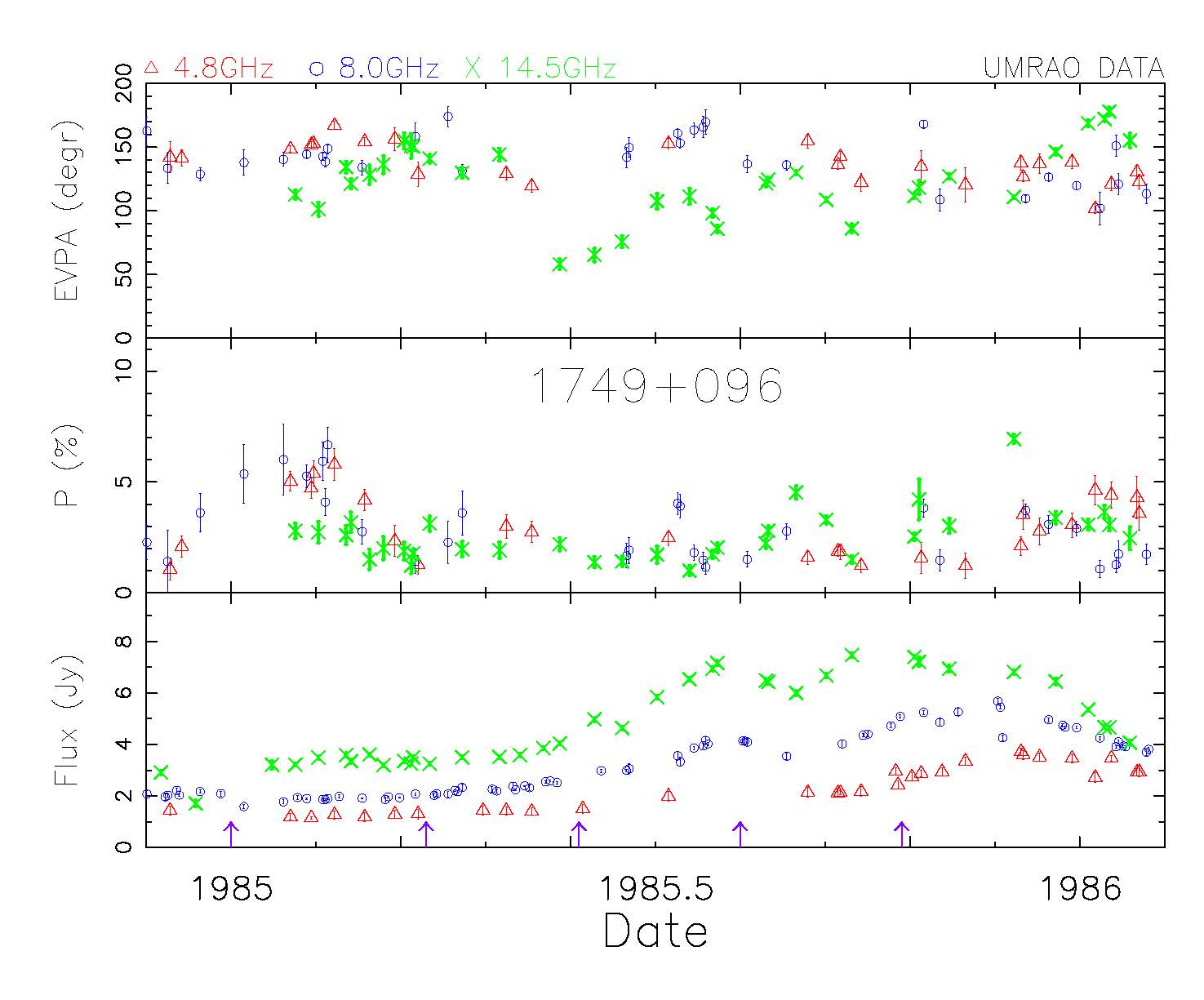} &
\hfill
  \includegraphics[width=3in]{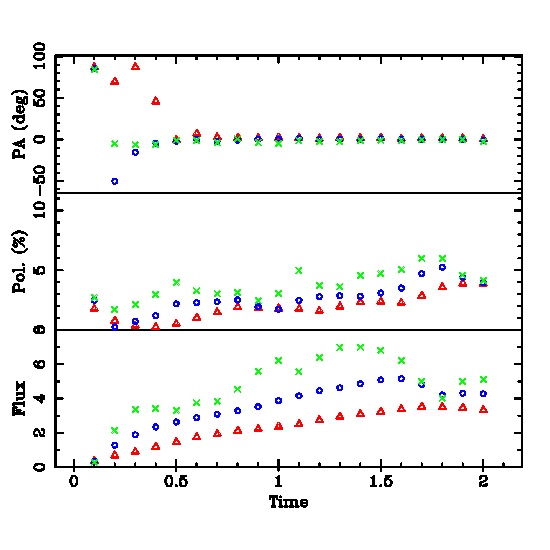}
\end{array}$
\end{center}

\caption{Left: The observed light curves during T1985 showing the UMRAO data at three frequencies during a series of outbursts in total flux density. The individual flares in the total flux light curve are blended, and they do not show the triangular shape noted for 0716+714. Upwards arrows along the abscissa mark the times of shock onset. Right: The simulated light curves using the adopted parameters.}
\end{figure}   

In Figure 3 we compare the data (left) and and the new simulation (right) for T1985.  The simulation is able to reproduce the spectral character and maximum amplitude of the total flux density (bottom panel), the detailed behavior of the fractional polarization light curve including the small flare near the start of the time window and the maximum value of P\% attained at all 3 frequencies, and the nearly constant value of the EVPA during most of the time window modeled. However, the details of the EVPA variability are not reproduced, and the variability apparent in the observed EVPA light curve is again very complex. 

In Figure 4 we show the data and the simulated light curves for T2008. Here we have again been able to reproduce the major features of the light curves including the spectral character and maximum amplitude of the total flux density, the low levels of the fractional polarization (under 5\% in general) and the fractional polarization spectral character during the activity, and the frequency-dependent EVPA separation. While we do not show the results for T2010, we have simulated the major features apparent in the UMRAO light curves using a model incorporating the parameters given in Table 3. The variability across each of these time windows is not self-similar, and scaling from a fiducial epoch in the simulations cannot reproduce the data at other epochs.

 \begin{figure}[H]
   \begin{center}$
\begin{array}{cc}
   \includegraphics[width =3.5in]{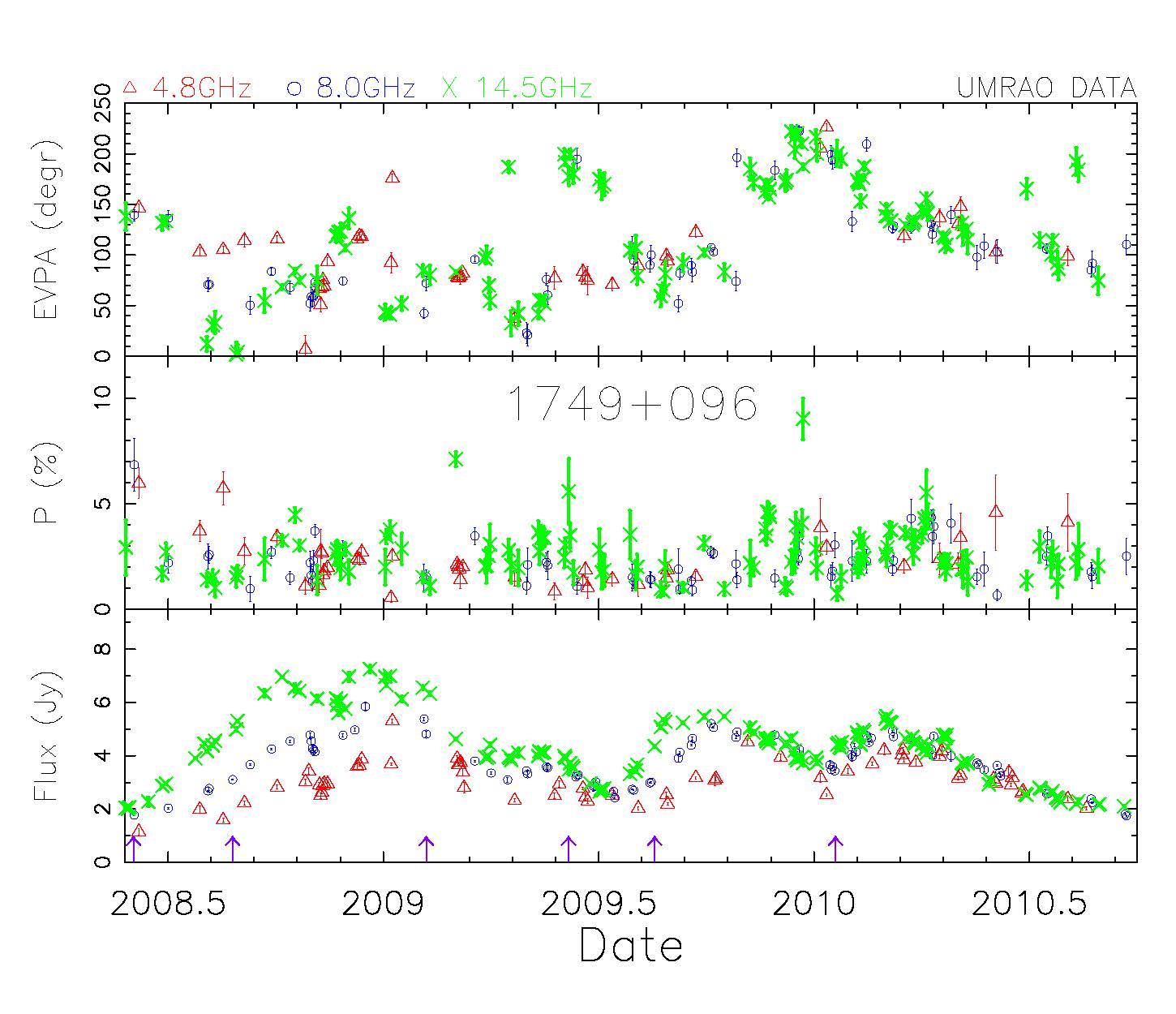} &
\hfill
  \includegraphics[width=3in]{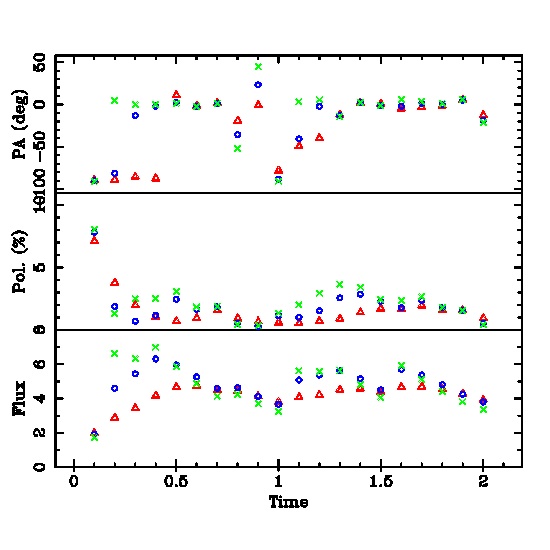}
\end{array}$
\end{center}
\caption{Left: The observed light curves during T2008. Upwards arrows along the abscissa mark the adopted shock onsets. Right: The simulated light curves using the adopted parameters listed in Table~3.}
\end{figure}

\begin{table}[H]
\caption{OT 081 Flow Parameters and Shock Characteristics}
\small 
\centering
\begin{tabular}{cccc}
\toprule
\textbf{Parameter}     & \textbf{T1985}	& \textbf{T2008}     &  \textbf{T2010}   \\
\midrule
Cutoff Lorentz Factor ($\gamma_i$)	& 50 	             &  50      & 50     \\
Bulk Lorentz Factor 	& 5		& 10                 &   10               \\
Viewing Angle        & 1.7$^{\circ}$    & 1.4$^{\circ}$      & 1.1$^{\circ}$    \\
Axial B field (B$_z$)  & 25\%           & 64\%               & 56\%               \\ 
Shock Sense            & F              &  F                 &  F                 \\
Number of Shocks       &  5             & 6                  &  3                 \\
Shock Obliquity        & 90$^{\circ}$   & 90$^{\circ}$       &    90$^{\circ}$      \\
Shock Compression      & 0.28 - 0.40    & 0.35 - 0.65        &   0.3 - 0.4         \\
Shock Lorentz Factor   &  14.5          &  24.7              &   27.3               \\
Shock $\beta_{app}$    &  10.4          &  21.8            &   22.5             \\

\bottomrule
\end{tabular}
\end{table}

Results for OT~081 spanning 1985 to 2012 are presented in Table 3. In addition to the parameters listed in Table 1 we include the shock Lorentz factor which comes from the bulk Lorentz factor of the quiescent flow and the shock strength (the compression factor). The shocks in this source are all moderately strong but not as strong as those in the 0716+714 time period modeled. The modeling identifies forward shocks, a viewing angle nearly in the line of sight, and component speeds in agreement with VLBA measurements at 15 and 22 GHz which are in the range 5 - 21 c based on VLBA data obtained over 11 years from 1995 to 2005 \cite{LU12}. While the flow speeds are consistent with VLBA estimates of the flow based on the maximum component speed, our low value of the viewing angle is not consistent with the value of 4.2$^{\circ}$ obtained from a combination of VLBA component speeds and a Doppler variability factor \cite{PYL09}. Unlike 0716+714, the low viewing angles obtained are very well constrained by the linear polarization data.

Comparison of the derived parameters as a function of time shows that systematic changes with time have occurred within the flow.  These are most extreme between T1985 and T2008, and the trends continue into T2010. As noted, these changes cannot be reproduced simply by scaling, and they are changes in the flow intrinsic to the source. However, from our analysis we cannot distinguish whether these are changes in the properties of the flow at a single location or whether the observer is looking at a different segment of the flow in each time window, possibly associated with changes in the orientation of the jet. The latter interpretation is plausible since changes in the orientation of the inner jet position angle have been identified from an analysis of MOJAVE data for this source  \cite{LIS13}. 

\subsection{Inclusion of a Helical Magnetic Field Component}
Because of the mounting observational evidence for the presence of helical magnetic fields in the parsec scale jets of blazars, including a 5-$\sigma$ determination of a helical/toroidal magnetic field in OT~081 \cite{GAB15}, we have investigated the effect of including a helical magnetic field on the simulated light curves for epoch T1985. This additional ordered component is included in the form of a magnetic flux rope which permeates the emitting plasma, and it is compressed along with the turbulent magnetic field component by the passage of a shock. While the  presence of a dominant helical magnetic field was ruled out in earlier work \cite{HAA11}, the new result presented here reflects the interplay of three magnetic field components.

\begin{figure}[H]
\begin{center}$
\begin{array}{cc}
   \includegraphics[width =3.0in]{OT_081_flare1.jpg} &
\hfill
  \includegraphics[width=3in]{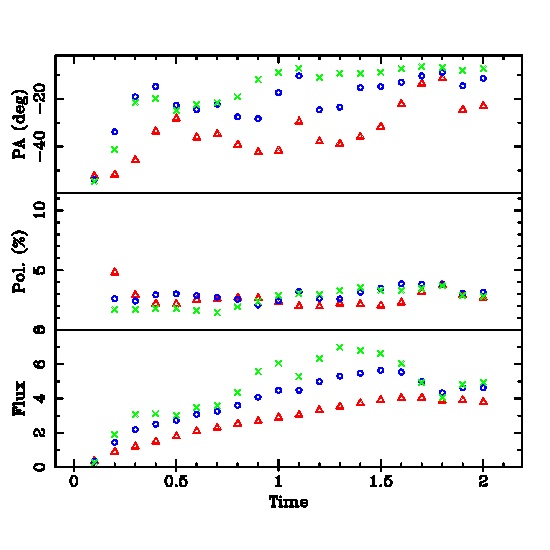}
\end{array}$
\end{center}
\caption{Left: The simulated light curves for the 1985.0 outburst based on the adopted parameters.  Right: The simulated light curves assuming the inclusion of an additional strong helical B field and the adopted values for all other parameters.}
\end{figure}   

The  light curve allowing for a turbulent magnetic field, an axial component, and the additional helical field component, is shown in Figure 5 right for the case of a strong helical component  (following  the notation of \cite{HAA11}, an order multiple $\geq$3).  This can be compared with the simulation based on the adopted parameters shown to the left. The effect of the inclusion of the helical field into the simulation is to suppress the  structure in the fractional linear polarization light curve and to produce a frequency-dependent separation of the  the EVPAs  with a spread by as much as $\Delta$EVPA=35$^\circ$. In contrast, the data exhibit a flat EVPA
spectrum which persists throughout most of the modeled time window. This simulation, thus, does not reproduce the major features of the observed fractional linear polarization and EVPA light curves. 

We conclude from a comparison of the simulation with the UMRAO data for  OT~081 that a dominant helical magnetic field is excluded. However, the inclusion of a modest ordered helical magnetic field component, in addition to the turbulent and axial components already described, improves the fit showing that the model can accommodate a  modest helical B field. This result supports a scenario where a modest helical magnetic field is present in the parsec scale jet downstream of the centimeter-band VLBI core region \cite{GAB14}. 

\section{Conclusions}
  We have shown that the spectral evolution of outbursts in total flux density and linear polarization at centimeter-band has successfully been simulated in blazars of both the BL Lac and QSO classes. Our analysis assumes a shock-in-jet model in which the radiation is produced within a particle dominated flow with an embedded, passive magnetic field which is primarily turbulent in character but in which ordered components also contribute to the energy density. A key point in the method is that the spectral variability in the linear polarization is the crucial constraint in the modeling, and that it is a particularly stringent constraint when the jet is oriented along or near to the line of sight to the observer. In contrast, the total flux density is nearly insensitive to changes in the key parameters. 

Using the modeling as a tool, we have identified both the jet flow and shock properties during selected outbursts in $\gamma$-ray-bright blazars. While simple analytic approaches can restrict the range of permitted values, detailed modeling such as we present here, is required to limit these ranges especially when extreme flow conditions may exist, such as in 0716+714 and in OT 081.   An additional advantage of the radiative transfer method is that it allows us to separate out a variety of processes. These include  relativistic effects such as Doppler boosts, and the signature on the emission of independently permitting a variety of magnetic field geometries: random, axial, and helical. Further, the long-term continuity of the UMRAO data, up to 4 decades for many sources, provides a method whereby intrinsic long term changes in blazar jets can be identified. This goal cannot be achieved using inhomogeneous data sets analyzed employing heterogeneous procedures. A knowledge of these parameters and their time-dependent behavior impacts our understanding of blazar emission including the processes responsible
for $\gamma$-ray flaring.  OT~081 has recently increased in photon flux by a factor of 20 in the GeV band over the average value during the period of our modeling \cite{GON16}, and this blazar has recently now been detected at TeV energies \cite{MAG16}. Modeling such as we demonstrate here can elucidate intrinsic changes in the flow associated with the onset of these high states in the High Energy and Very High Energy bands.

We are continuing to refine the shock-in-jet model described here to improve the agreement between the archival UMRAO data and the simulations. Including a modest contribution from an ordered helical magnetic field in OT~081 has already  demonstrated that future inclusion may improve the fits in other outbursts modeled, and the importance of this additional magnetic field component, which the simulations suggest contributes only modestly to the overall energy density of the system, needs to be explored more fully. Additionally we would like to allow for a range of shock obliquities since our current models do not reproduce the details of the observed EVPA spectral evolution. Our modeling assumes that the shocks are non-interacting and moving at constant velocity in rectilinear motion, while there is evidence from VLBI measurements that all of these assumptions are violated and that must also affect to some degree the interpretation of the  variability in the light curves. Nevertheless, the modeling has already proven to be an effective tool in investigating the intrinsic properties of blazar jets and in identifying extreme conditions.

\vspace{6pt} 

%
\acknowledgments{Acquisition of the UMRAO data and model development were supported by a series of NASA Fermi GI grants (NNX09AU16G, NNX10AP16G, NNX11AO13G, and NNX13AP18G), and by a series of grants from the NSF (most recently AST 0607523). The operation of UMRAO was supported in part by the University of Michigan. Computational resources and services were provided by Advanced Research Computing at the University of Michigan, Ann Arbor. T.H. was supported in part by an award from the Jenny and Antti Wihuri foundation and by the Academy of Finland project number 267324. This work has made used of data from the MOJAVE website which is maintained by the MOJAVE team.}

\conflictofinterests{The authors declare no conflict of interest.}

\renewcommand\bibname{References}

\end{document}